\def \oot {\frac{1}{2}}
\def \Jp {$J/\psi$ }
\def\pr{{^\prime}}

\pdfoutput=1
\documentclass[preprint,preprintnumbers,amsmath,amssymb]{revtex4}
 
\usepackage{graphicx}% Include figure files
\usepackage{epstopdf}
\usepackage{dcolumn}% Align table columns on decimal point
\usepackage{bm}% bold math

\begin{document}

\title{Quarkonium above deconfinement as an open quantum system}
\author{C. Young}
\email{cyoung@grad.physics.sunysb.edu}
\affiliation{Department of Physics \& Astronomy, State University of New York, Stony Brook, NY 11794-3800, U.S.A.}
\author{K. Dusling}
\email{kdusling@quark.phy.bnl.gov}
\affiliation{Physics Department, Building 510A\\Brookhaven National Laboratory\\Upton, NY-11973, USA}

\date{\today}

\begin{abstract}

Quarkonium at temperatures above deconfinement is modeled as an open quantum system, 
whose dynamics is determined not just by a potential energy and mass, but also by a drag coefficient 
which characterizes its interaction with the medium. The reduced density matrix for a heavy 
particle experiencing dissipative forces is expressed as an integral over paths in imaginary time and evaluated numerically. We demonstrate 
that dissipation could affect the Euclidean heavy-heavy correlators calculated in lattice simulations at 
temperatures just above deconfinement.

\end{abstract}

\maketitle

\section{Introduction}

Because of the masses of heavy quarks relative to the QCD energy scale $\Lambda_{QCD}$, heavy-heavy bound states are systems 
which can be described to some accuracy with first quantization techniques \cite{Appelquist:1974yr, Eichten:1974af, Eichten:1978tg}. 
At zero temperature, lattice calculations of a heavy-heavy system give a potential term which can 
describe quarkonium spectroscopy. The potential has a non-trivial temperature dependence. For this reason, it has been the hope of both theorists and experimentalists for over twenty years that changes in the yields of quarkonium in heavy-ion collisions would be a signal for changes in the temperature and phase of the medium \cite{Matsui:1986dk}. Since then, 
analysis of heavy-ion experiments have shown that some suppression of \Jp yields at the RHIC is anomalous \cite{Adare:2006ns}. However, there is little agreement among theorists over the cause of this suppression. 
It is clear that the dynamics of quarkonium above the deconfinement phase transition should be considered carefully. 

One step in this direction was made by Shuryak and one of us, who modeled charmonium in heavy-ion collisions as undergoing 
Brownian motion, with the heavy quark diffusion coefficient $D_H$ taken to be quite small, as expected from phenomenology of single heavy quarks \cite{Adare:2006nq, Moore:2004tg} and by gauge-gravity duality \cite{CasalderreySolana:2006rq, Gubser:2006bz, Herzog:2006gh}. This work 
\cite{Young:2008he}
demonstrated that the survival of \Jp states in plasma cannot be determined just by examining 
whether or not the temperature-dependent potentials allow bound states; instead, the dynamics of charm and charmonium are determined by multiple interactions with the medium, and the time scales 
of the collision are very important.

Another important step in understanding the dynamics of 
charmonium is provided by newer lattice calculations \cite{Jakovac:2006sf}. In these calculations, the Euclidean correlators for local operators related to quarkonium 
spectroscopy are calculated, namely, $G(\tau,T) = \int d^3x \left \langle J_H({\bf x},\tau) J_H({\bf 0}, 0) 
\right \rangle$, where $J_H({\bf x}, \tau)=\bar{\psi}({\bf x},\tau) \Gamma \psi({\bf x},\tau)$ is a local 
composite operator related to a given mesonic channel. Taking the results of these lattice simulations, 
Mocsy and Petreczky compared these correlators, at various temperatures and in different 
channels, with the results of two different potential models \cite{Mocsy:2007bk}. 
In this work, there is a low-frequency contribution to the spectral densities used to determine the 
Euclidean correlator, which is due to the diffusion of heavy quarks \cite{Petreczky:2005nh}, however, for 
$\omega\gg T^2/M$, the diffusion of heavy quarks is assumed not to have any effect on the dynamics of 
quarkonium.

In this paper, we will outline how the imaginary-time propagator with periodicity $\beta=1/T$ can 
be determined for quarkonium systems interacting with a heat bath. In Section \ref{TPIF}, we
review the reduced density matrix, apply the model of Caldeira and Leggett for quantum Brownian 
motion to imaginary time, and then describe briefly path-integral Monte Carlo techniques for calculating this reduced density matrix. In Section \ref{QAAO}, we show how the imaginary-time propagator for 
charmonium may be calculated to high accuracy with these techniques, for any potential or drag 
coefficient, and we show the results for $G(\tau, T)$ for the vector channel at temperatures just above 
the deconfinement transition.

Not many before have considered the effect of the medium on heavy quark and quarkonium Euclidean 
correlators calculated on the lattice. Current work by Beraudo et al. \cite{Beraudo:2009zz} does consider the 
medium effect on heavy quark correlators. Starting from the QCD Lagrangian, they make the HTL 
approximation for the heavy quark's interaction with the light degrees of freedom at finite temperature, 
and then they integrate these degrees of freedom out. The focus of our work is on a heavy particle 
interacting quite strongly with the medium, as suggested by calculations taking advantage of gauge-gravity duality \cite{CasalderreySolana:2006rq}. We show how the Lagrangian for a particle interacting strongly with a heat bath can be renormalized so that functional integration is well-defined, and where analytic results are possible.

\section{The path integral form for the reduced density matrix of a dissipative system}
\label{TPIF}

For an excellent review of the functional integral approach to quantum Brownian motion, in both 
imaginary and real time, see \cite{Grabert:1988yt}. For a briefer discussion of these ideas and how they apply to quarkonium, see \cite{Young:INTTalk2009}. These results are discussed in generality, and 
shown to arise from the Schwinger-Keldysh contour integral for a heavy quark's real-time partition function, by Son and Teaney \cite{Son:2009vu}. These systems can be studied with approaches besides the path-integral formulation, and the study of quantum Brownian motion, in terms of a partial differential equation for the density matrix, has been studied by Hu, Paz, and Zhang \cite{Hu:1991di, Hu:1993vs}.

\subsection{The reduced density matrix}

Without loss of generality, consider a system consisting of a heavy particle of mass $M$ which we will call the system, 
minimally coupled to a harmonic oscillator of mass $m$.
\begin{eqnarray}
L & = & L_S+L_I {\rm ;} \nonumber \\
L_S & = & \oot  M\dot{x}^2-V(x){\rm ,} \nonumber \\
L_I & = & \oot  m\dot{r}^2-\oot  m\omega^2r^2-Cxr .
\label{L}
\end{eqnarray}
We analytically continue this Lagrangian to $\tau=it$.
\begin{eqnarray}
S^E_S[x]&=&\int_0^\beta\left[\oot  M\dot{x}^2+V(x)\right]{\rm ,} \nonumber \\
S^E_I[x,r] &=&\int_0^{\beta}\left[\oot m\dot{r}^2+\oot m\omega^2
  r^2+Cxr\right]d\tau
\label{S^E_I}
\end{eqnarray}
We can simplify this Lagrangian by making a change of variables.  We subtract the 
particular solution to the classical equations of motion as determined by Eq. \ref{S^E_I}.
\begin{eqnarray}
r(\tau) & \equiv & r\pr(\tau)+\frac{C}{m\omega}\int_0^{\tau}
d\tau\pr x(\tau\pr)\sinh\left[\omega(\tau-\tau\pr)\right] \nonumber \\
 & \equiv & r\pr(\tau)+A[x,\tau] 
\end{eqnarray}
In terms of the shifted coordinate $r\pr$ the Euclidean action becomes that of a simple harmonic oscillator:
\begin{eqnarray}
S^E_I[x,r] & = & \int_0^{\beta}\left[ \oot m\dot{r}^{\prime 2}+\oot 
m\omega^2r^{\prime 2}+\oot Cx(\tau)A[x,\tau]\right]d\tau \nonumber \\
 & & +m\dot{A}[x,\beta]\left(r\pr(\beta)+\oot A[x,\beta]\right) \nonumber \\
 & \equiv & S^{\prime E}_I[x,r\pr] {\rm .}
 \label{S^E_simple}
\end{eqnarray}

As always, the propagator of a system for imaginary time $\beta=-i/T$ gives matrix 
elements of the thermal density operator. In our example, the density matrix has 4 indices, 2 for 
each degree of freedom of the system.   In the example given above the density matrix is given as
\begin{eqnarray}
\rho(x_i,r_i;x_f,r_f;\beta)  =  \int_{x(0)=x_i}^{x(\beta)=x_f} {\cal D}x \int_{r(0)=r_i}^{r(\beta)=r_f}{\cal D}r
\exp\left(-S^E_S[x]-S^E_I[x,r]\right){\rm .}
\end{eqnarray}
If we were never interested in measurements of the 
degree of freedom $r$, we could take the trace over the indices corresponding to this degree of 
freedom, and work with a density operator with only two indices. With this in mind, we 
define the {\it reduced density matrix}:
\begin{equation}
\rho_{red}(x_i,x_f,\beta) \equiv \int dr \rho(x_i,r;x_f,r;\beta){\rm .}
\end{equation}
In the next section it will be the only operator 
practical for calculating thermal averages.

For the system defined by Eq.~\ref{L}, we can now write a path-integral description of the reduced 
density matrix,
\begin{eqnarray}
\rho_{red}(x_i,x_f,\beta)  & = & \int dr \int {\cal D}x \;{\cal D}r\pr
\;\exp\left(-S^E_S[x]-S^E_I[x,r\pr]\right) \nonumber \\
 & = & \int {\cal D}x \exp\left(-S^E_x[x]\right)\int dr \int {\cal D}r\pr
\exp\left(-S^{'E}_I[x,r\pr]\right) {\rm ,} 
\label{rho_red_path}
\end{eqnarray}
where we integrate over paths with endpoints
$x(0)=x_i$, $x(\beta)=x_f$, $r\pr(0)=r$, and $r\pr(\beta)=r-A[x,\beta]$. The final step in 
\ref{rho_red_path} allows the integral over the paths $r\pr(\tau)$ to be performed independently 
of the integral over $x(\tau)$. Using Eq. \ref{S^E_simple}, a simple Gaussian integral yields
\begin{eqnarray}
&&\rho_{red}(x_i,x_f,\beta) =\int_{x(0)=x_i}^{x(\beta)=x_f} {\cal D}x \nonumber\\ &\times&\exp \left(-S^E_S[x]+\sum_i\frac{C^2_i}{2m\omega_i \sinh(\frac{\omega_i\beta}{2})}
 \int_0^\beta d\tau \int_0^{\tau}ds\; x(\tau)x(s)\cosh\left[\omega_i\left(\tau-s-\beta/2\right)\right] \right) {\rm ,}
 \label{rho_red_fin}
\end{eqnarray}
where the summation is introduced because we have generalized this to a system where a heavy particle interacts with a bath of simple harmonic oscillators.
This is now the path integral form for the reduced density matrix. This is analogous to the idea of influence functionals, worked out for real-time propagators many years 
ago \cite{Feynman:1965}.

\subsection{Making a dissipative system}

Any finite quantum-mechanical system is reversible and therefore inappropriate for describing 
Brownian motion. One might find it intuitive that if the bath of harmonic oscillators were taken to 
an infinite limit, it would be ``large enough" so that energy from the heavy particle could dissipate 
into the system and never return. This intuition was proven to be true by the authors of 
\cite{Caldeira:1982iu}, 
who considered the real-time evolution of the density matrix for our system and showed that 
when the bath of harmonic oscillators is determined by the continuous density of states
\begin{equation}
C^2(\omega)\rho_D(\omega) = \begin{cases} 
\frac{2m\eta\omega^2}{\pi}  & \text{if } \omega < \Omega\\ 
0 & \text{if } \omega > \Omega \end{cases}
\end{equation}
the force autocorrelator for the heavy particle is proportional to $\delta(t-t^{'})$ at high temperatures. 
In this ``white noise" limit, the density matrix describes an ensemble of particles interacting according 
to the Langevin equation, which has been used to describe Brownian motion for a long time. It is a 
stochastic differential equation, is irreversible, and evolves any ensemble towards the thermal 
phase space distribution.

We now take this density of states and substitute it into Eq.~\ref{rho_red_fin}
\begin{eqnarray}
&&\rho_{red}(x_i,x_f,\beta) =\int_{x(0)=x_i}^{x(\beta)=x_f} {\cal D}x \nonumber\\ &\times&\exp \left(-S^E_S[x]  +\frac{\eta}{\pi}\int_0^{\Omega}d\omega
\int_0^\beta d\tau \int_0^{\tau}ds \;x(\tau)x(s)\frac{\omega \cosh\left[\omega(\tau-s-\beta/2)\right]}{\sinh(\frac{\omega \beta}{2})}\right) {\rm .} 
 \label{rho_red_fin2}
\end{eqnarray}

The divergences of this action can be isolated by integrating by parts twice
\begin{eqnarray}
&&\int_0^{\Omega}d\omega
\int_0^\beta d\tau \int_0^{\tau}ds \;x(\tau)x(s)\frac{\omega \cosh\left[\omega(\tau-s-\beta/2)\right]}{\sinh(\frac{\omega \beta}{2})} \nonumber \\
 & =&\Omega \int_0^{\beta} d\tau\; (x(\tau))^2
-\oot (x_i-x_f)^2 \ln(M\Omega/\eta)
\frac{\cosh(\Omega\beta/2)-1}{\sinh(\Omega\beta/2)}\nonumber \\
 & & -\frac{1}{2} (x_i-x_f)^2\left[\gamma_E+\ln\left(\frac{\eta\beta}{\pi M}\right)\right]\nonumber \\
 & & +\left(x_i-x_f\right)  \int_0^{\beta}d\tau\; \dot{x}(\tau)\;{\rm lnsin}\left(\frac{\pi\tau}{\beta}\right) \nonumber \\
 & & +\int_0^{\beta}d\tau \int_0^{\tau}ds\;
\dot{x}(\tau)\dot{x}(s)\;{\rm lnsin}\left(\frac{\pi(\tau-s)}{\beta}\right), 
\end{eqnarray}
where ${\rm lnsin}(x)\equiv \ln\left[\sin\left(x\right)\right]$ and $\gamma_E$ is the Euler-Mascheroni constant.  The first two terms on the right-hand side correspond to a renormalization of the potential for 
the heavy particle, always necessary when considering the interaction of a particle with infinitely 
many additional degrees of freedom. They are temperature-independent in the limit of large 
$\Omega$, and may be introduced as temperature-independent modifications of the Lagrangian for the particle. The final three terms are finite and well-behaved, and have been evaluated in the limit 
$\Omega \to \infty$. The final form for the reduced density matrix becomes
\begin{eqnarray}
\rho_{red}(x_i,x_f,\beta) =\int_{x(0)=x_i}^{x(\beta)=x_f} {\cal D}x \;\exp\bigg\{&-&S^E_S[x]\nonumber\\ 
 &-&\frac{\eta}{2\pi} (x_i-x_f)^2\left[\gamma_E+\ln\left(\frac{\eta\beta}{\pi M}\right)\right]\nonumber\\
  &+&\frac{\eta}{\pi}\left(x_i-x_f\right)  \int_0^{\beta}d\tau\; \dot{x}(\tau)\;{\rm lnsin}\left(\frac{\pi\tau}{\beta}\right)\nonumber\\ 
  &+&\frac{\eta}{\pi}\int_0^{\beta}d\tau \int_0^{\tau}ds\;
\dot{x}(\tau)\dot{x}(s)\;{\rm lnsin}\left(\frac{\pi(\tau-s)}{\beta}\right)
\bigg\}.
 \label{rho_red_final}
\end{eqnarray}

In summary, we have determined an expression for the reduced density matrix of a system (consisting of a massive particle in a potential) interacting with a bath of oscillators.  The coupling to the bath has been chosen in order to reproduce the results of classical Brownian motion in the high temperature limit. We have taken care to make the term in the exponential finite, by isolating the divergences through integration by parts. This is important for path integral Monte Carlo simulation to be possible for this functional integral.

\subsection{Example: The otherwise free particle}
\label{free}

The reduced density matrix for a particle interacting with such a bath can be determined analytically for the otherwise free particle ($V_R(x)=0$). First, write an arbitrary path as an expansion around the classical solution
\begin{eqnarray}
x(\tau) & = & x_{cl}(\tau)+\xi(\tau){\rm ;} \nonumber \\
x_{cl}(\tau) & \equiv & x_i+(x_f-x_i)\tau/\beta{\rm ,}
\nonumber \\
\xi(\tau) & \equiv & \sum_{n=1}^{\infty}c_n\sin\left(\frac{n\pi \tau}{\beta}\right){\rm .} 
\end{eqnarray}
One can find an analytic solution by substituting this into Eq.~\ref{rho_red_final} .  We skip the details: after evaluating numerous integrals using contour integration, and then changing variables for the 
integration over the even Fourier coefficients, we find the reduced density matrix
\begin{eqnarray}
\rho_{red}(x_i,x_f,\beta) & = & \sqrt{\frac{M}{2\pi\beta}+\frac{\eta}{2\pi^2}\left[\log(2)+\gamma_E+\Psi\left(1+\frac{\eta\beta}{2\pi M}\right)\right]}
\nonumber \\
 &  \times& \exp\left\{-\left[\frac{M}{2\beta}+\frac{\eta}{2\pi}\left[\log\left(\frac{\eta\beta}{2\pi M}\right)
-\Psi\left(1+\frac{\eta\beta}{2\pi M}\right)\right] \right]\left(x_i-x_f\right)^2\right\}{\rm ,} 
\end{eqnarray}
where $\Psi(x)$ is the digamma function and the 
overall normalization is determined by analytic continuation of $\beta$ and requiring   the 
propagator to conserve probability for purely imaginary $\beta$.  One interesting physical result is easily obtained from the Fourier transform of the reduced density 
matrix
\begin{equation}
\left \langle p^2 \right \rangle = \frac{M_{eff}}{\beta}{\rm ,}\;\;\; M_{eff}=M+\frac{\eta \beta}{\pi}
\left[\log\left(\frac{\eta \beta}{2\pi M}\right)-\Psi\left(1+\frac{\eta \beta}{2\pi M}\right) \right]{\rm .}
\label{meff}
\end{equation}

\subsection{The path-integral Monte Carlo algorithm for the reduced density matrix}

Obtaining the analytic result for our reduced density matrix was reasonable for the otherwise 
free particle. For the simple harmonic oscillator, the analytic result exists but has a rather complicated 
expression. For the potentials which describe quarkonium spectroscopy with some precision, analytic 
work becomes entirely impractical. We would like to use numerical simulation to obtain 
reliable estimates of reduced density matrices and Euclidean correlators.

The most natural numerical approach for the formalism we adopted is path-integral Monte Carlo 
(PIMC). For an excellent review of the technique, see the review of D. M. Ceperley \cite{ceperley}. 
Paths are either sampled according to a Metropolis algorithm determined by the action of 
interest, or sampled from some convenient distribution which samples the entire space of paths 
with some weight. For our work with a single degree of freedom, we found that sampling a convenient 
distribution (in our case, the distribution of paths determined by $\exp(-S_{free}[x])$) to be 
sufficient, which is easily sampled for any discretization of the path with a bisection method. When sampling the space of paths, the next step is to determine an estimate for the action of each path. 
For our case, the primitive action with the simplest integration of the new dissipative terms is 
sufficient. Once this is determined, any correlator can be calculated by sampling paths and 
making weighted averages.

\section{Quarkonium as an open quantum system}
\label{QAAO}

We should now apply this functional integral to the problem of quarkonium at temperatures above 
deconfinement. To this end, let us rework the Euclidean correlators calculated on the lattice to a form 
with which we can calculate.

For a given channel, the two-point Euclidean correlator for a composite mesonic operator is given by 
\begin{eqnarray}
G(\tau,T)  &= & \int d^3 x \left \langle J_{\Omega}({\bf x},\tau)J_{\Omega}({\bf 0}, 0) 
\right \rangle_{\beta}{\rm ,} \nonumber \\
J_{\Omega}({\bf x},\tau) & = & \bar{\psi}({\bf x},\tau) \Omega \psi({\bf x}, \tau) {\rm ,}
\end{eqnarray}
where $\Omega=1{\rm ,}\; \gamma^0{\rm ,}\; \gamma^{\mu}{\rm ,}\; \gamma^0\gamma^{\mu}$ determine 
the mesonic channel to be scalar, pseudo-scalar, vector, or pseudo-vector, respectively. 

For now, consider only the vector channel (the following arguments must be modified for the scalar 
and pseudo-vector channels). In order to 
switch to a first-quantization approach for the energy range where a potential model would be 
appropriate, think of this correlator as being the sum of the expectation values of an operator over 
all of the states in the Fock space of $N$-particle mesonic systems, with each expectation value 
entering the sum weighted by the state's Boltzmann factor. One term in this sum, of course, is the ``vacuum" expectation value
\begin{equation}
\label{vev}
\left \langle 0|J^{\mu}({\bf x},\tau) J_{\mu}({\bf 0},0)|0 \right \rangle
= \left \langle \mu, {\bf x}; \tau | \mu, {\bf 0}; 0 \right \rangle_{\rm light} {\rm ,}
\end{equation}
where the subscript represents that the trace is still taken over all of the light degrees of freedom in 
QCD.

The right-hand side of Equation \ref{vev} represents an important step: by decomposing the mesonic 
operators into products of creation and annihilation operators, the vacuum expectation value can be 
related to the overlap of two one-particle states. Also, note here that since expectation values of the higher states enter into the thermal average multiplied by factors which are roughly powers of 
$\exp(-2M_Q/T)$, the vacuum expectation value dominates the thermal average in the limit $M\gg T$.
Therefore, we have identified the leading contribution to the correlator in the vector channel.

The trace over the light degrees of freedom is of course the central problem of QCD, and no analytic result exists. However, in the infinite mass limit, this has been computed on the lattice as the expectation value of two Polyakov loops \cite{Kaczmarek}. The dissipative effects on this propagator have been studied, as we noted previously, by gauge-gravity duality. Our ansatz here is that these results describe well the dynamics of sufficiently heavy quarks.

The path integral we wish to evaluate can be written in terms of relative and absolute coordinates as
\begin{eqnarray}
\label{PI_rel}
\left \langle {\bf x}, {\bf x}; \tau \vert {\bf 0}, {\bf 0}; 0 \right \rangle 
= \int {\cal D}{\bf X} &\exp&\left\{ -\int_0^{\beta}d\tau M\dot{{\bf X}}^2
+\frac{2\eta}{\pi}\int_0^\beta d\tau \int_0^{\tau}ds\;\dot{{\bf X}}(\tau)\dot{{\bf X}}(s)\;{\rm lnsin}\left(\frac{\pi}{\beta}(\tau-s)\right) \right\}\nonumber \\ 
\times  \int {\cal D} {\bf x}_{rel}&\exp&\left\{-\int_0^{\beta}d\tau \left[\frac{1}{4} M \dot{{\bf x}}_{rel}^2 
+V({\bf x}_{rel}) \right]\right.\nonumber \\
 &&\;\;\;+ \left.\frac{\eta}{2\pi}\int_0^\beta d\tau \int_0^{\tau}ds\;\dot{{\bf x}}_{rel}(\tau)\dot{{\bf x}}_{rel}(s)\;{\rm lnsin}\left(\frac{\pi}{\beta}(\tau-s)\right) \right\}{\rm ,}
\end{eqnarray}
where the functional integration are over all paths satisfying ${\bf X}(0)={\bf 0}$, ${\bf X}(\tau)={\bf x}$, 
${\bf X}(\beta)={\bf 0}$, ${\bf x}_{rel}(0)={\bf 0}$, ${\bf x}_{rel}(\tau)={\bf 0}$, and 
${\bf x}_{rel}(\beta)={\bf 0}$.

The right-hand side of Equation \ref{PI_rel} can now be calculated, for any potential, with a PIMC 
simulation. We use the Cornell potential for the interaction between the heavy quarks.

\begin{figure}
\includegraphics[height=9cm]{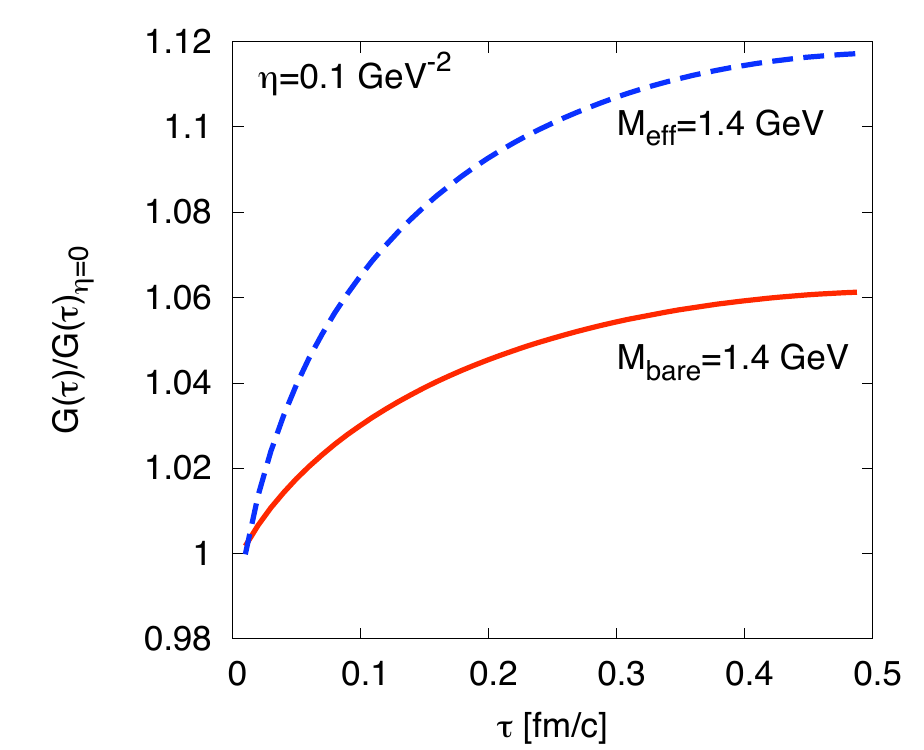}
\caption{
$G(\tau)/G_{\eta=0}$ evaluated for $\eta=0.1$ ${\rm GeV}^{-2}$ at $T=1.2 T_c$.
 }
\label{gtau}
\end{figure}

To obtain $G(\tau)$, the first term on the right-hand side of Eq.~\ref{PI_rel} is integrated with respect to ${\bf X}$, yielding the diagonal value for the free-particle reduced density matrix which is independent of 
$\tau$.   In Figure \ref{gtau} we show the mesonic correlator as a function of $\tau$ at $T=1.2 T_c$.  We have chosen a representative value of $\eta =0.1$ GeV$^{-2}$ and show results for fixed bare and effective charm mass.

Clearly, dissipative effects due to the medium change the mesonic correlators measured in lattice 
simulations. Also, these effects are not trivial: holding the effective charm mass fixed but varying $\eta$ still 
leads to changes in the mesonic correlators. The next step is to perform spectral analysis of these 
correlators to examine how dissipation affects the spectral function for this correlator. From here, the dissipative effects can be included in heavy-ion phenomenology.

\section{Conclusions}
\label{conclusions}

We have taken the approach of Caldeira and Leggett and determined an imaginary-time path integral, 
representing the reduced density matrix for a dissipative system, which contains integrals that are convergent and therefore suitable for numerical integration. We have shown how 
interactions with the QCD medium above deconfinement can affect Euclidean heavy-heavy 
correlators.

The purpose of this paper is only to identify this effect.  In order to achieve a result which is appropriate 
for comparisons with the lattice data, the following work must be done: 1.) the mass of the heavy 
quark must be changed whenever the drag coefficient is changed, since we are concerned with results 
for $G(\tau)$ with different values of $\eta$ but the same ``effective" mass, 2.) the maximum entropy 
method must be applied to $G(\tau)$, so that the spectral function corresponding to the correlation 
function will be extracted with high precision, 3.) this spectral function must have the zero-mode 
contribution added and be cut so that the perturbative result for the spectral function is used at 
large $\omega$ instead, and finally 4.) this spectral function should be used to calculate $G(\tau)$.
However, let us note here that this work can be put to greater use in examining quarkonium in heavy-ion collisions. 

Let us argue one last time for treating quarkonium above deconfinement as an open quantum system:
typically in particle physics, the rate for the scattering of an $N$-particle state into another is 
determined 
from the square of the matrix element whose indices are the initial and final states, where these 
states are in the momentum basis. This makes perfect sense in high-energy experiments, where 
the incoming and outgoing states are basically each 2-particle momentum eigenstates and the matrix 
element can be expanded in terms of a small coupling. Such an approach, however, seems entirely 
inappropriate for quarkonium formed in a heavy-ion collision, whose constituent quarks are localized 
in position relative to the surrounding medium and are strongly coupled both to the medium and each other. Non-perturbative techniques and experiment both suggest that quarkonium rapidly thermalizes 
and interacts strongly with the medium. The most reasonable approach for explaining the observables 
is to describe quarkonium with a reduced density matrix, whose evolution is determined by a potential 
$and$ a drag coefficient which are both determined non-perturbatively.

\section{Acknowledgments}

We would like to thank E. Shuryak and D. Teaney for very useful and detailed suggestions concerning our work.  
CY thanks the Institute for Nuclear Theory at the University of Washington for its 
hospitality during the completion of this work.  CY was supported by the US-DOE grants 
DE-FG02-88ER40388 and DE-FG03-97ER4014.  KD was supported by the US-DOE grant DE-AC02-98CH10886.


\begin{thebibliography}{99}

\bibitem{Appelquist:1974yr}
  T.~Appelquist, A.~De Rujula, H.~D.~Politzer and S.~L.~Glashow,
  %``Charmonium Spectroscopy,''
  Phys.\ Rev.\ Lett.\  {\bf 34}, 365 (1975).
  %%CITATION = PRLTA,34,365;%%

\bibitem{Eichten:1974af}
  E.~Eichten, K.~Gottfried, T.~Kinoshita, J.~B.~Kogut, K.~D.~Lane and T.~M.~Yan,
  %``The Spectrum Of Charmonium,''
  Phys.\ Rev.\ Lett.\  {\bf 34}, 369 (1975)
  [Erratum-ibid.\  {\bf 36}, 1276 (1976)].
  %%CITATION = PRLTA,34,369;%%

\bibitem{Eichten:1978tg}
  E.~Eichten, K.~Gottfried, T.~Kinoshita, K.~D.~Lane and T.~M.~Yan,
  %``Charmonium: The Model,''
  Phys.\ Rev.\  D {\bf 17}, 3090 (1978)
  [Erratum-ibid.\  D {\bf 21}, 313 (1980)].
  %%CITATION = PHRVA,D17,3090;%%

\bibitem{Matsui:1986dk}
  T.~Matsui and H.~Satz,
  %``J/psi Suppression by Quark-Gluon Plasma Formation,''
  Phys.\ Lett.\  B {\bf 178}, 416 (1986).
  %%CITATION = PHLTA,B178,416;%%

\bibitem{Adare:2006ns}
  A.~Adare {\it et al.}  [PHENIX Collaboration],
  %``J/psi production vs centrality, transverse momentum, and rapidity in Au  +
  %Au collisions at s(NN)**(1/2) = 200-GeV,''
  Phys.\ Rev.\ Lett.\  {\bf 98}, 232301 (2007)
  [arXiv:nucl-ex/0611020].
  %%CITATION = PRLTA,98,232301;%%

\bibitem{Jakovac:2006sf}
  A.~Jakovac, P.~Petreczky, K.~Petrov and A.~Velytsky,
  %``Quarkonium correlators and spectral functions at zero and finite
  %temperature,''
  Phys.\ Rev.\  D {\bf 75}, 014506 (2007)
  [arXiv:hep-lat/0611017].
  %%CITATION = PHRVA,D75,014506;%%

\bibitem{Young:2008he}
  C.~Young and E.~Shuryak,
  %``Charmonium in strongly coupled quark-gluon plasma,''
  Phys.\ Rev.\  C {\bf 79}, 034907 (2009)
  [arXiv:0803.2866 [nucl-th]].
  %%CITATION = PHRVA,C79,034907;%%
 
\bibitem{Adare:2006nq}
  A.~Adare {\it et al.}  [PHENIX Collaboration],
  %``Energy Loss and Flow of Heavy Quarks in Au+Au Collisions at sqrt(s_NN) =
  %200 GeV,''
  Phys.\ Rev.\ Lett.\  {\bf 98}, 172301 (2007)
  [arXiv:nucl-ex/0611018].
  %%CITATION = PRLTA,98,172301;%% 
  
  %\cite{Moore:2004tg}
\bibitem{Moore:2004tg}
  G.~D.~Moore and D.~Teaney,
  %``How much do heavy quarks thermalize in a heavy ion collision?,''
  Phys.\ Rev.\  C {\bf 71}, 064904 (2005)
  [arXiv:hep-ph/0412346].
  %%CITATION = PHRVA,C71,064904;%%
  
  %\cite{CasalderreySolana:2006rq}
\bibitem{CasalderreySolana:2006rq}
  J.~Casalderrey-Solana and D.~Teaney,
  %``Heavy quark diffusion in strongly coupled N = 4 Yang Mills,''
  Phys.\ Rev.\  D {\bf 74}, 085012 (2006)
  [arXiv:hep-ph/0605199].
  %%CITATION = PHRVA,D74,085012;%%

  %\cite{Gubser:2006bz}
\bibitem{Gubser:2006bz}
  S.~S.~Gubser,
  %``Drag force in AdS/CFT,''
  Phys.\ Rev.\  D {\bf 74}, 126005 (2006)
  [arXiv:hep-th/0605182].
  %%CITATION = PHRVA,D74,126005;%%
  
  %\cite{Herzog:2006gh}
\bibitem{Herzog:2006gh}
  C.~P.~Herzog, A.~Karch, P.~Kovtun, C.~Kozcaz and L.~G.~Yaffe,
  %``Energy loss of a heavy quark moving through N = 4 supersymmetric
  %Yang-Mills plasma,''
  JHEP {\bf 0607}, 013 (2006)
  [arXiv:hep-th/0605158].
  %%CITATION = JHEPA,0607,013;%%

\bibitem{Mocsy:2007bk}
  A.~Mocsy and P.~Petreczky,
  %``Quarkonium correlators at finite temperature and potential models,''
  PoS {\bf LAT2007}, 216 (2007)
  [arXiv:0710.5205 [hep-lat]].
  %%CITATION = POSCI,LAT2007,216;%%

\bibitem{Petreczky:2005nh}
  P.~Petreczky and D.~Teaney,
  %``Heavy quark diffusion from the lattice,''
  Phys.\ Rev.\  D {\bf 73}, 014508 (2006)
  [arXiv:hep-ph/0507318].
  %%CITATION = PHRVA,D73,014508;%%

\bibitem{Beraudo:2009zz}
  A.~Beraudo, J.~P.~Blaizot, G.~Garberoglio and P.~Faccioli,
  %``Heavy-quarks in the QGP: study of medium effects through euclidean
  %propagators and spectral functions,''
  Nucl.\ Phys.\  A {\bf 830}, 319C (2009)
  [arXiv:0907.1797 [hep-ph]].
  %%CITATION = NUPHA,A830,319C;%%

\bibitem{Grabert:1988yt}
  H.~Grabert, P.~Schramm and G.~L.~Ingold,
  %``Quantum Brownian motion: The Functional inegral approach,''
  Phys.\ Rept.\  {\bf 168}, 115 (1988).
  %%CITATION = PRPLC,168,115;%%

\bibitem{Young:INTTalk2009}
 C.~Young at the INT-CATHIE Joint mini-program ``Quarkonium in hot media: from QCD to Experiment," June 22, 2009

\bibitem{Son:2009vu}
  D.~T.~Son and D.~Teaney,
  %``Thermal Noise and Stochastic Strings in AdS/CFT,''
  JHEP {\bf 0907}, 021 (2009)
  [arXiv:0901.2338 [hep-th]].
  %%CITATION = JHEPA,0907,021;%%

\bibitem{Hu:1991di}
 B.~L.~Hu, J.~P.~Paz and Y.~h.~Zhang,
 %``Quantum Brownian motion in a general environment: 1. Exact master equation
 %with nonlocal dissipation and colored noise,''
 Phys.\ Rev.\  D {\bf 45}, 2843 (1992).
 %%CITATION = PHRVA,D45,2843;%%

\bibitem{Hu:1993vs}
 B.~L.~Hu, J.~P.~Paz and Y.~Zhang,
 %``Quantum Brownian motion in a general environment. 2: Nonlinear coupling and
 %perturbative approach,''
 Phys.\ Rev.\  D {\bf 47}, 1576 (1993).
 %%CITATION = PHRVA,D47,1576;%%

\bibitem{Feynman:1965}
R.P. Feynman and A.R. Hibbs, {\it Quantum Mechanics and Path Integrals} (McGraw-Hill, New York, 
1965)

\bibitem{Caldeira:1982iu}
  A.~O.~Caldeira and A.~J.~Leggett,
  %``Path integral approach to quantum Brownian motion,''
  Physica {\bf 121A}, 587 (1983).
  %%CITATION = PHYSA,121A,587;%%

\bibitem{ceperley} D. M. Ceperley, {\em Reviews of Modern Physics} Vol. 67, 
No. 2, April 1995

\bibitem{Kaczmarek}
  O.~Kaczmarek, F.~Karsch, P.~Petreczky and F.~Zantow,
  %``Heavy Quark Anti-Quark Free Energy and the Renormalized Polyakov Loop,''
  Phys.\ Lett.\  B {\bf 543}, 41 (2002)
  [arXiv:hep-lat/0207002].
  %%CITATION = PHLTA,B543,41;%%

\end{thebibliography}
\end{document}